\def\year{2019}\relax
\documentclass[letterpaper]{article} 
\usepackage{aaai19}  
\usepackage[utf8]{inputenc} 
\usepackage[T1]{fontenc}    
\usepackage{times}  
\usepackage{helvet}  
\usepackage{courier}  
\usepackage{url}  
\usepackage{graphicx}  
\usepackage{svg}
\usepackage{array}
\usepackage{subfig}
\newcolumntype{L}[1]{>{\raggedright\let\newline\\\arraybackslash\hspace{0pt}}m{#1}}

\begin{document}
%
\title{Hierarchical Macro Strategy Model for \\ MOBA Game AI}
\author{$^{1}$Bin Wu, $^{1}$Qiang Fu, $^{1}$Jing Liang, $^{1}$Peng Qu, $^{1}$Xiaoqian Li, $^{1}$Liang Wang, \AND $^{2}$Wei Liu, $^{1}$Wei Yang, $^{1}$Yongsheng Liu\\
$^{1,2}$Tencent AI Lab\\
$^{1}$\{benbinwu, leonfu, masonliang, pengqu, xiaoqianli, enginewang, willyang, kakarliu\}@tencent.com\\$^{2}$wliu@ee.columbia.edu\\
}
\maketitle
\begin{abstract}
The next challenge of game AI lies in Real Time Strategy (RTS) games. RTS games provide partially observable gaming environments, where agents interact with one another in an action space much larger than that of GO. Mastering RTS games requires both strong macro strategies and delicate micro level execution. Recently, great progress has been made in micro level execution, while complete solutions for macro strategies are still lacking. In this paper, we propose a novel learning-based Hierarchical Macro Strategy model for mastering MOBA games, a sub-genre of RTS games. Trained by the Hierarchical Macro Strategy model, agents explicitly make macro strategy decisions and further guide their micro level execution. Moreover, each of the agents makes independent strategy decisions, while simultaneously communicating with the allies through leveraging a novel imitated cross-agent communication mechanism. We perform comprehensive evaluations on a popular 5v5 Multiplayer Online Battle Arena (MOBA) game. Our 5-AI team achieves a 48\% winning rate against human player teams which are ranked top 1\% in the player ranking system.
\end{abstract}

\section{Introduction}

Light has been shed on artificial general intelligence after AlphaGo defeated world GO champion Lee Seedol \cite{silver2016mastering}. Since then, game AI has drawn unprecedented attention from not only researchers but also the public. Game AI aims much more than robots playing games. Rather, games provide ideal environments that simulate the real world. AI researchers can conduct experiments in games, and transfer successful AI ability to the real world. 

Although AlphaGo is a milestone to the goal of general AI, the class of problems it represents is still simple compared to the real world. Therefore, recently researchers have put much attention to real time strategy (RTS) games such as Defense of the Ancients (Dota) \cite{dota} and StarCraft \cite{vinyals2017starcraft,tian2017elf}, which represents a class of problems with next level complexity. Dota is a famous set of science fiction 5v5 Multiplayer Online Battle Arena (MOBA) games. Each player controls one unit and cooperate with four allies to defend allies' turrets, attack enemies' turrets, collect resources by killing creeps, etc. The goal is to destroy enemies' base. 



There are four major aspects that make RTS games much more difficult compared to GO: 1) \textbf{Computational complexity}. The computational complexity in terms of action space or state space of RTS games can be up to $10^{20,000}$, while the complexity of GO is about $10^{250}$  \cite{openaifive0}. 2) \textbf{Multi-agent}. Playing RTS games usually involves multiple agents. It is crucial for multiple agents to coordinate and cooporate. 3) \textbf{Imperfect information}. Different to GO, many RTS games make use of fog of war \cite{vinyals2017starcraft} to increase game uncertainty. When the game map is not fully observable, it is essential to consider gaming among one another. 4) \textbf{Sparse and delayed rewards}. Learning upon game rewards in GO is challenging because the rewards are usually sparse and delayed. RTS game length could often be larger than 20,000 frames, while each GO game is usually no more than 361 steps.

To master RTS games, players need to have strong skills in both macro strategy operation and micro level execution. In recent study, much attention and attempts have been put to micro level execution \cite{vinyals2017starcraft,tian2017elf,synnaeve2011bayesian,wender2012applying}. So far, Dota2 AI developed by OpenAI using reinforcement learning, i.e., OpenAI Five, has made the most advanced progress \cite{dota}. OpenAI Five was trained directly on micro level action space using proximal policy optimization algorithms along with team rewards \cite{schulman2017proximal}. OpenAI Five has shown strong teamfights skills and coordination comparable to top professional Dota2 teams during a demonstration match held in The International 2018 \cite{ti8}. OpenAI's approach did not explicitly model macro strategy and tried to learn the entire game using micro level play. However, OpenAI Five was not able to defeat professional teams due to weakness in macro strategy management \cite{openaifive,openaifive2}. 

Related work has also been done in explicit macro strategy operation, mostly focused on navigation. Navigation aims to provide reasonable destination spots and efficient routes for agents. Most related work in navigation used influence maps or potential fields \cite{deloura2001game,hagelback2008rise,do2015development}. Influence maps quantify units using handcrafted equations. Then, multiple influence maps are fused using rules to provide a single-value output to navigate agents. Providing destination is the most important purpose of navigation in terms of macro strategy operation. The ability to get to the right spots at right time makes essential difference between high level players and the others. Planning has also been used in macro strategy operation. Ontanon \textit{et al}. proposed Adversarial Hierarchical-Task Network (AHTN) Planning \cite{ontanon2015adversarial} to search hierarchical tasks in RTS game playing. Although AHTN shows promising results in a mini-RTS game, it suffers from efficiency issue which makes it difficult to apply to full MOBA games directly.


Despite of the rich and promising literature, previous work in macro strategy failed to provide complete solution:

First, reasoning macro strategy implicitly by learning upon micro level action space may be too difficult. OpenAI Five's ability gap between micro level execution and macro strategy operation was obvious. It might be over-optimistic to leave models to figure out high level strategies by simply looking at micro level actions and rewards. We consider explicit macro strategy level modeling to be necessary. 

Second, previous work on explicit macro strategy heavily relied on handcrafted equations for influence maps/potential fields computation and fusion. In practice, there are usually thousands of numerical parameters to manually decide, which makes it nearly impossible to achieve good performance. Planning methods on the other hand cannot meet efficiency requirement of full MOBA games.

Third, one of the most challenging problems in RTS game macro strategy operation is coordination among multiple agents. Nevertheless, to the best of our knowledge, previous work did not consider it in an explicit way. OpenAI Five considers multi-agent coordination using team rewards on micro level modeling. However, each agent of OpenAI Five makes decision without being aware of allies' macro strategy decisions, making it difficult to develop top coordination ability in macro strategy level. 

Finally, we have found that modeling strategic phase is crucial for MOBA game AI performance. However, to the best of our knowledge, previous work did not consider this.

Teaching agents to learn macro strategy operation, however, is challenging. Mathematically defining macro strategy, e.g., besiege and split push, is difficult in the first place. Also, incorporating macro strategy on top of OpenAI Five's reinforcement learning framework \cite{dota} requires corresponding execution to gain rewards, while macro strategy execution is a complex ability to learn by itself. Therefore, we consider supervised learning to be a better scheme because high quality game replays can be fully leveraged to learn macro strategy along with corresponding execution samples. Note that macro strategy and execution learned using supervised learning can further act as an initial policy for reinforcement learning.

In this paper, we propose Hierarchical Macro Strategy (HMS) model - a general supervised learning framework for MOBA games such as Dota. HMS directly tackles with \textbf{computational complexity} and \textbf{multi-agent} challenges of MOBA games. More specifically, HMS is a hierarchical model which conducts macro strategy operation by predicting attention on the game map under guidance of game phase modeling. Thereby, HMS reduces computational complexity by incorporating game knowledge. Moreover, each HMS agent conducts learning with a novel mechanism of communication with teammates agents to cope with multi-agent challenge. Finally, we have conducted extensive experiments in a popular MOBA game to evaluate our AI ability. We matched with hundreds of human player teams that ranked above 99\% of players in the ranked system and achieved 48\% winning rate.



The rest of this paper is organized as follows: First, we briefly introduce Multiplayer Online Battle Arena (MOBA) games and compare the computational complexity with GO. Second, we illustrate our proposed Hierarchical Macro Strategy model. Then, we present experimental results in the fourth section. Finally, we conclude and discuss future work.

\section{Multiplayer Online Battle Arena (MOBA) Games}

\subsection{Game Description}

MOBA is currently the most popular sub-genre of the RTS games. MOBA games are responsible for more than 30\% of the online gameplay all over the world, with titles such as Dota, League of Legends, and Honour of Kings \cite{moba_hot}. According to a worldwide digital games market report in February 2018, MOBA games ranked first in grossing in both PC and mobile games \cite{hok}. 

In MOBA, the standard game mode requires two 5-player teams play against each other. Each player controls one unit, i.e., hero. There are numerous of heroes in MOBA, e.g., more than 80 in Honour of Kings. Each hero is uniquely designed with special characteristics and skills. Players control movement and skill releasing of heroes via the game interface.


\begin{figure*}[]
	\centering
	\subfloat[]{{{\label{ui}\includegraphics[width=1\columnwidth]{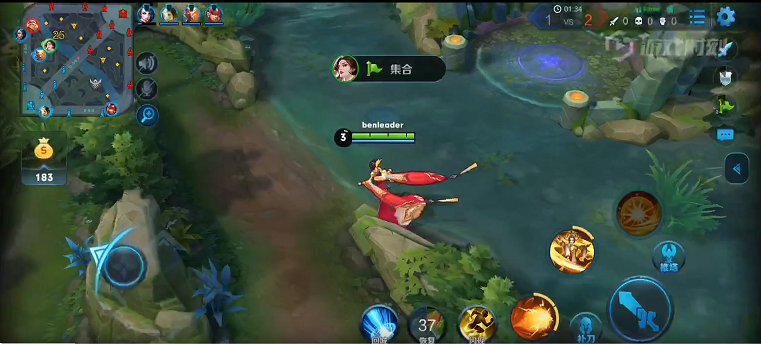}}}}
	\subfloat[]{{{\label{map}\includegraphics[width=0.75\columnwidth]{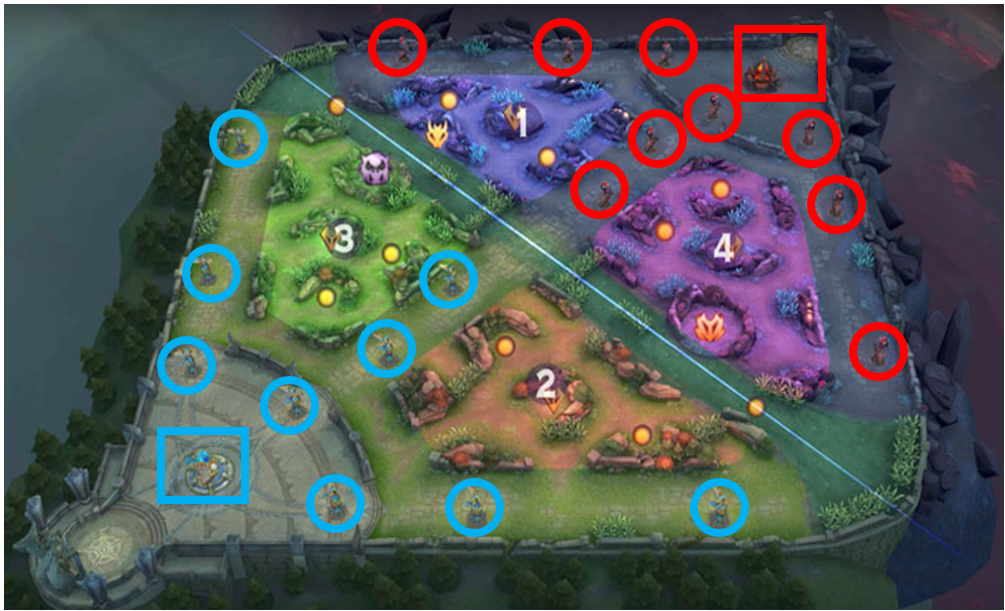}}}}
	\caption{(a) Game UI of Honour of Kings. Players use left bottom steer button to control movements, while right bottom set of buttons to control skills. Players can observe surroundings via the screen and view the mini full map using the left top corner. (b) An example map of MOBA. The two teams are colored in blue and red, each possesses nine turrets (circled in rounds) and one base (circled in squares). The four jungle areas are numbered from 1 to 4.}
	\label{moba}
\end{figure*}


As shown in Figure. \ref{ui}, Honour of Kings players use left bottom steer button to control movements, while right bottom set of buttons to control skills. Surroundings are observable via the main screen. Players can also learn full map situation via the left top corner mini-map, where observable turrets, creeps, and heroes are displayed as thumbnails. Units are only observable either if they are allies' units or if they are within a certain distance to allies' units.

There are three lanes of turrets for each team to defend, three turrets in each lane. There are also four jungle areas on the map, where creep resources can be collected to increase gold and experience. Each hero starts with minimum gold and level 1. Each team tries to leverage resources to obtain as much gold and experience as possible to purchase items and upgrade levels. The final goal is to destroy enemy's base. A conceptual map of MOBA is shown in Figure. \ref{map}.


To master MOBA games, players need to have both excellent macro strategy operation and proficient micro level execution. Common macro strategies consist of opening, laning, ganking, ambushing, etc. Proficient micro level execution requires high accuracy of control and deep understanding of damage and effects of skills. Both macro strategy operation and micro level execution require mastery of timing to excel, which makes it extremely challenging and interesting. More discussion of MOBA can be found in \cite{silva2017moba}. 

Next, we will quantify the computational complexity of MOBA using Honour of Kings as an example.

\subsection{Computational Complexity}

The normal game length of Honour of Kings is about 20 minutes, i.e., approximately 20,000 frames in terms of gamecore. At each frame, players make decision with tens of options, including movement button with 24 directions, and a few skill buttons with corresponding releasing position/directions. Even with significant discretization and simplification, as well as reaction time increased to 200ms, the action space is at magnitude of $10^{1,500}$.

As for state space, the resolution of Honour of Kings map is 130,000 by 130,000 pixels, and the diameter of each unit is 1,000. At each frame, each unit may have different status such as hit points, levels, gold. Again, the state space is at magnitude of $10^{20,000}$ with significant simplification.

Comparison of action space and state space between MOBA and GO is listed in Table. \ref{tab:comparison}.

\begin{table}[]
	\centering
	
	\caption{Computational complexity comparison between GO and MOBA.}
	\begin{tabular}{|L{1.5cm}|L{3cm}|L{3cm}|}
		\hline
		&	GO		&	MOBA\\
		\hline
		Action Space	&	$250^{150} \approx 10^{360}$ (250 pos available, 150 decisions per game in average) & $10^{1500}$ (10 options, 1500 actions per game)\\
		\hline
		State Space		&	$3^{360} \approx 10^{170}$ (361 pos, 3 states each)	&	$10^{20000}$ (10 heroes, 2000+pos * 10+states)\\
		\hline
	\end{tabular}
	\label{tab:comparison}
\end{table}

\subsection{MOBA AI Macro Strategy Architecture}

Our motivation of designing MOBA AI macro strategy model was inspired from how human players make strategic decisions. During MOBA games, experienced human players are fully aware of game phases, e.g., opening phase, laning phase, mid game phase, and late game phase \cite{silva2017moba}. During each phase, players pay attention to the game map and make corresponding decision on where to dispatch the heroes. For example, during the laning phase players tend to focus more on their own lanes rather than backing up allies, while during mid to late phases, players pay more attention to teamfight spots and pushing enemies' base. 

To sum up, we formulate the macro strategy operation process as "phase recognition -> attention prediction -> execution". To model this process, we propose a two-layer macro strategy architecture, i.e., phase and attention:

\begin{itemize}
	\item \textbf{Phase} layer aims to recognize current game phase so that attention layer can have better sense about where to pay attention to. \\
	\item \textbf{Attention} layer aims to predict the best region on game maps to dispatch heroes. \\
\end{itemize}

\textbf{Phase} and \textbf{Attention} layers act as high level guidance for micro level execution. We will describe details of modeling in the next section. The network structure of micro level model is almost identical to the one used in OpenAI Five\footnote{\url{https://d4mucfpksywv.cloudfront.net/research-covers/openai-five/network-architecture.pdf}} \cite{dota}, but in a supervised learning manner. We did minor modification to adapt it to Honour of Kings, such as deleting Teleport.



\section{Hierarchical Macro Strategy Model}

We propose a Hierarchical Macro Strategy (HMS) model to consider both phase and attention layers in a unified neural network. We will first present the unified network architecture. Then, we illustrate how we construct each of the phase and attention layers.

\subsection{Model Overview}

We propose a Hierarchical Macro Strategy model (HMS) to model both attention and phase layers as a multi-task model. It takes game features as input. The output consists of two tasks, i.e., attention layer as the main task and phase layer as an auxiliary task. The output of attention layer directly conveys macro strategy embedding to micro level models, while resource layer acts as an axillary task which help refine the shared layers between attention and phase tasks. 

\begin{figure*}[ht]
	\centering
	\includegraphics[width=1.7\columnwidth]{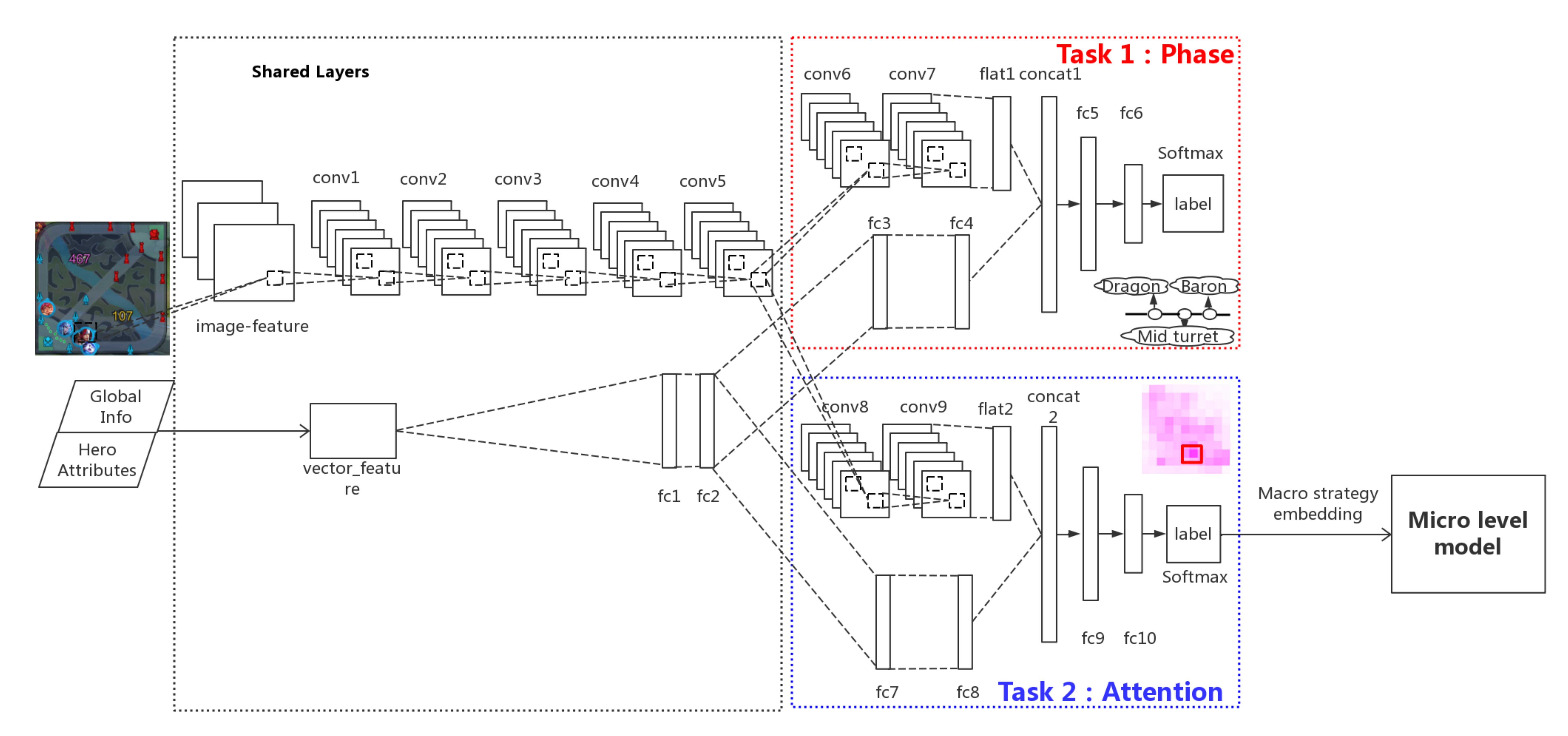}
	\caption{Network Architecture of Hierarchical Macro Strategy Model}
	\label{hms_network}
\end{figure*}

The illustrating network structure of HMS is listed in Figure. \ref{hms_network}. HMS takes both image and vector features as input, carrying visual features and global features respectively. In image part, we use convolutional layers. In vector part, we use fully connected layers. The image and vector parts merge in two separate tasks, i.e., attention and phase. Ultimately, attention and phase tasks take input from shared layers through their own layers and output to compute loss. 

\subsection{Attention Layer}

\begin{figure}[]
	\centering
	\includegraphics[width=1\columnwidth]{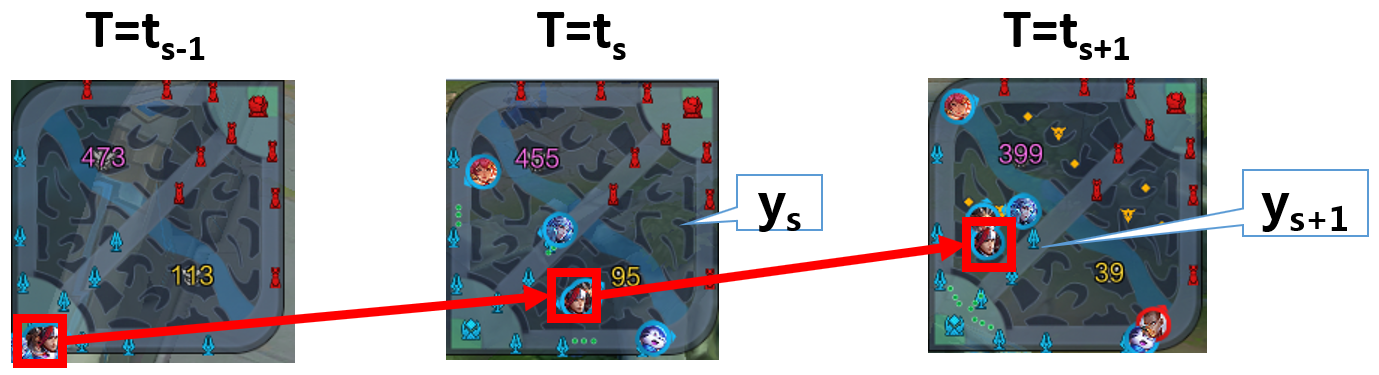}
	\caption{Illustrating example for label extraction in attention layer.}
	\label{transfer_label}
\end{figure}

Similar to how players make decisions according to the game map, attention layer predicts the best region for agents to move to. However, it is tricky to tell from data that where is a player's destination. We observe that regions where attack takes place can be indicator of players' destination, because otherwise players would not have spent time on such spots. According to this observation, we define ground-truth regions as the regions where players conduct their next attack. An illustrating example is shown in Figure. \ref{transfer_label}.

Let $s$ to be one session in a game which contains several frames, and $s-1$ indicates the session right before $s$. In Figure. \ref{transfer_label}, $s-1$ is the first session in the game. Let $t_s$ to be the starting frame of $s$. Note that a session ends along with attack behavior, therefore there exists a region $y_s$ in $t_s$ where the hero conducts attack. As shown in Figure. \ref{transfer_label}, label for $s-1$ is $y_s$, while label for $s$ is $y_{s+1}$. Intuitively, by setting up labels in this way, we expect agents to learn to move to $y_s$ at the beginning of game. Similarly, agents are supposed to move to appropriate regions given game situation.

\subsection{Phase layer}


Phase layer aims to recognize the current phase. Extracting game phases ground-truth is difficult because phase definition used by human players is abstract. Although roughly correlated to time, phases such as opening, laning, and late game depend on complicated judgment based on current game situation, which makes it difficult to extract ground-truth of game phases from replays. Fortunately, we observe clear correlation between game phases with major resources. For example, during the opening phase players usually aim at taking outer turrets and baron, while for late game, players operate to destroy enemies' base.

\begin{figure*}[]
	\centering
	\subfloat[]{{{\label{strategy_layer}\includegraphics[width=0.5\columnwidth]{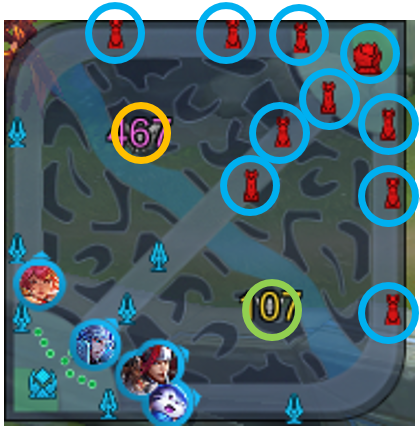}}}}
	\subfloat[]{{{\label{strategy_series}\includegraphics[width=1.2\columnwidth]{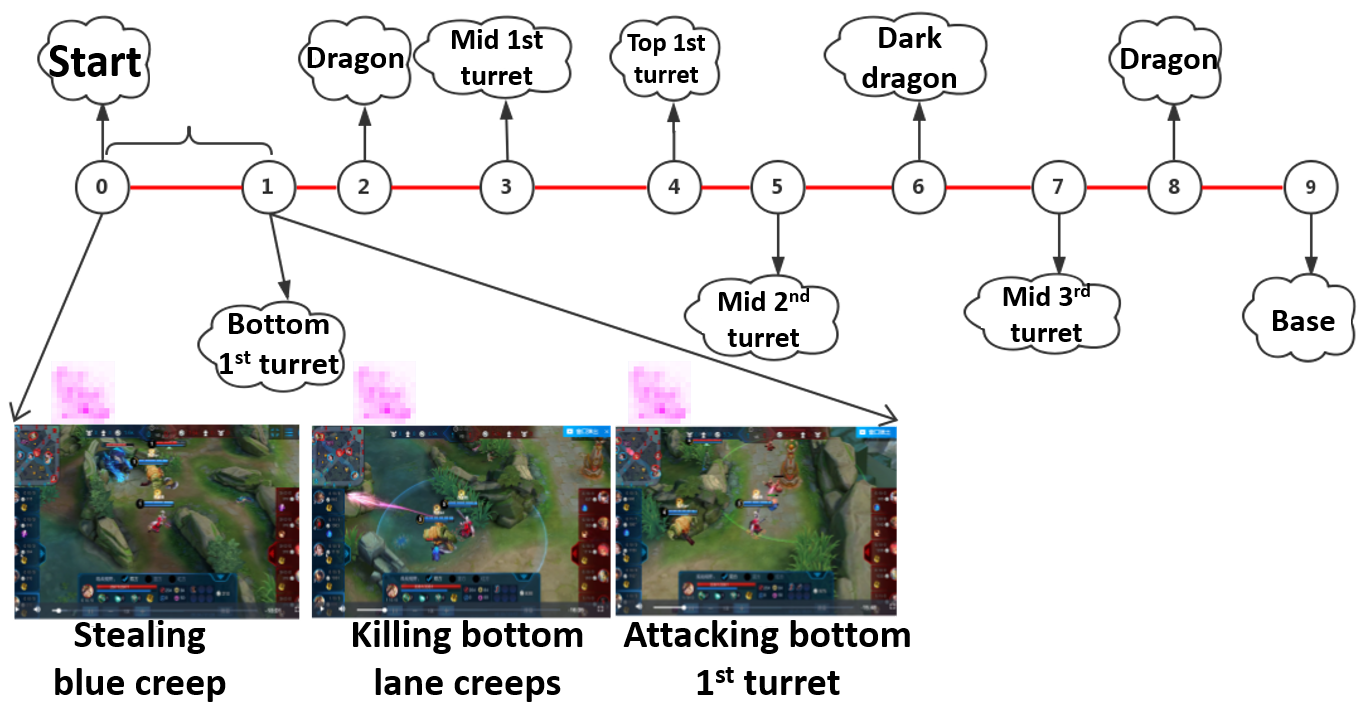}}}}
	\caption{(a) Major resources (circled, i.e., turrets, base, dragon, and baron) modeled in phase layer. (b) Illustrating example for label extraction in phase layer.}
	\label{strategy_model}
\end{figure*}

Therefore, we propose to model phases with respect to major resources. More specifically, major resources indicate turrets, baron, dragon, and base. We marked the major resources on the map in Figure. \ref{strategy_layer}. Label definition of phase layer is similar to attention layer. The only difference is that $y_s$ in phase layer indicates attack behavior on turrets, baron, dragon, and base instead of in regions. Intuitively, phase layer modeling splits the entire game into several phases via modeling which macro resource to take in current phase. 

We do not consider other resources such as lane creeps, heroes, and neutral creeps as major objectives because usually these resources are for bigger goal, such as destroying turrets or base with higher chance. Figure. \ref{strategy_series} shows a series of attack behavior during the bottom outer turret strategy. The player killed two neutral creeps in the nearby jungle and several lane creeps in the bottom lane before attacking the bottom outer turret.


We expect the model to learn when and what major resources to take given game situation, and in the meanwhile learn attention distribution that serve each of the major resources.

\subsection{Imitated Cross-agents Communication}

Cross-agents communication is essential for a team of agents to cooperate. There is rich literature of cross-agent communication on multi-agent reinforcement learning research \cite{sukhbaatar2016learning,foerster2016learning}. However, it is challenging to learn communication using training data in supervised learning because the actual communication is unknown. 

To enable agents to communicate in supervised learning setting, we have designed a novel cross-agents communication mechanism. During training phase, we put attention labels of allies as features for training. During testing phase, we put attention prediction of allies as features and make decision correspondingly. In this way, our agents can "communicate" with one another and learn to cooperate upon allies' decisions. We name this mechanism as Imitated Cross-agents Communication due to its supervised nature.

\section{Experiments}

In this section, we evaluate our model performance. We first describe the experimental setup, including data preparation and model setup. Then, we present qualitative results such as attention distribution under different phase. Finally, we list the statistics of matches with human player teams and evaluate improvement brought by our proposed model. 

\subsection{Experimental Setup}

\subsubsection{Data Preparation}

To train a model, we collect around 300 thousand game replays made of King Professional League competition and training records. Finally, 250 million instances were used for training. We consider both visual and attributes features. On visual side, we extract 85 features such as position and hit points of all units and then blur the visual features into 12*12 resolution. On attributes side, we extract 181 features such as roles of heroes, time period of game, hero ID, heroes' gold and level status and Kill-Death-Assistance statistics.

\subsubsection{Model Setup}

We use a mixture of convolutional and fully-connected layers to take inputs from visual and attributes features respectively. On convolutional side, we set five shared convolutional layers, each with 512 channels, $padding=1$, and one RELU. Each of the tasks has two convolutional layers with exactly the same configuration with shared layers. On fully-connected layers side, we set two shared fully-connected layers with 512 nodes. Each of the tasks has two fully-connected layers with exactly the same configuration with shared layers. Then, we use one concatenation layer and two fully-connected layers to fuse results of convolutional layers and fully-connected layers. We use ADAM as the optimizer with base learning rate at 10e-6. Batch size was set at 128. The loss weights of both phase and attention tasks are set at 1. We used CAFFE \cite{jia2014caffe} with eight GPU cards. The duration to train an HMS model was about 12 hours. 

Finally, the output for attention layer corresponds to 144 regions of the map, resolution of which is exactly the same as the visual inputs. The output of the phase task corresponds to 14 major resources circled in Figure. \ref{strategy_layer}. 

%
%
%




\subsection{Experimental Results}

\subsubsection{Opening Attention}

\begin{figure*}[]
	\centering
	\includegraphics[width=2\columnwidth]{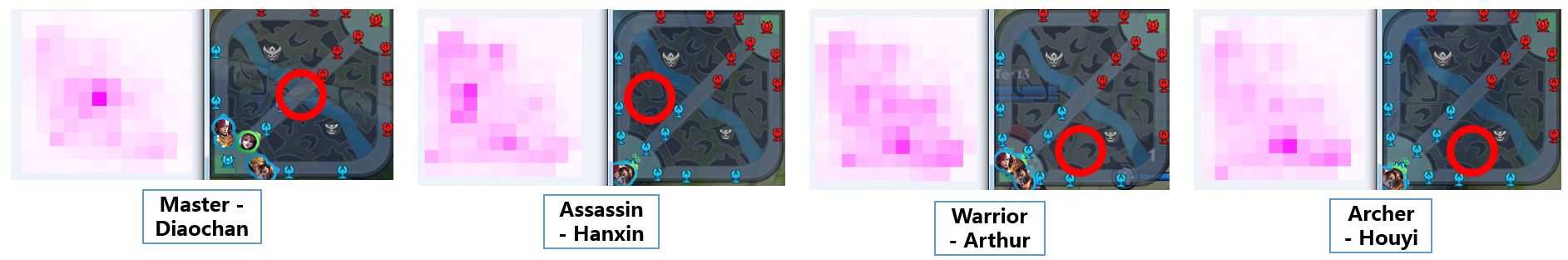}
	\caption{One of the opening strategies learned for different hero roles. The hottest regions are highlighted with red circle.}
	\label{opening}
\end{figure*}

Opening is one of the most important strategies in MOBA. We show one opening attention of different heroes learned by our model in Figure. \ref{opening}. In Figure. \ref{opening}, each subfigure consists of two square images. The left-hand-side square image indicates the attention distribution of the right-hand-side MOBA mini-map. The hottest region is highlighted with red circle. We list attention prediction of four heroes, i.e., Diaochan, Hanxin, Arthur, and Houyi. The four heroes belong to master, assasin, warrior, and archer respectively. According to the attention prediction, Diaochan is dispatched to middle lane, Hanxin will move to left jungle area, and Authur and Houyi will guard the bottom jungle area. The fifth hero Miyamoto Musashi, which was not plotted, will guard the top outer turret. This opening is considered safe and efficient, and widely used in Honour of Kings games.

\subsubsection{Attention Distribution Affected by Phase Layer}

\begin{figure}[h]
	\centering
	\subfloat[]{{{\label{upper_outer}\includegraphics[width=0.49\columnwidth]{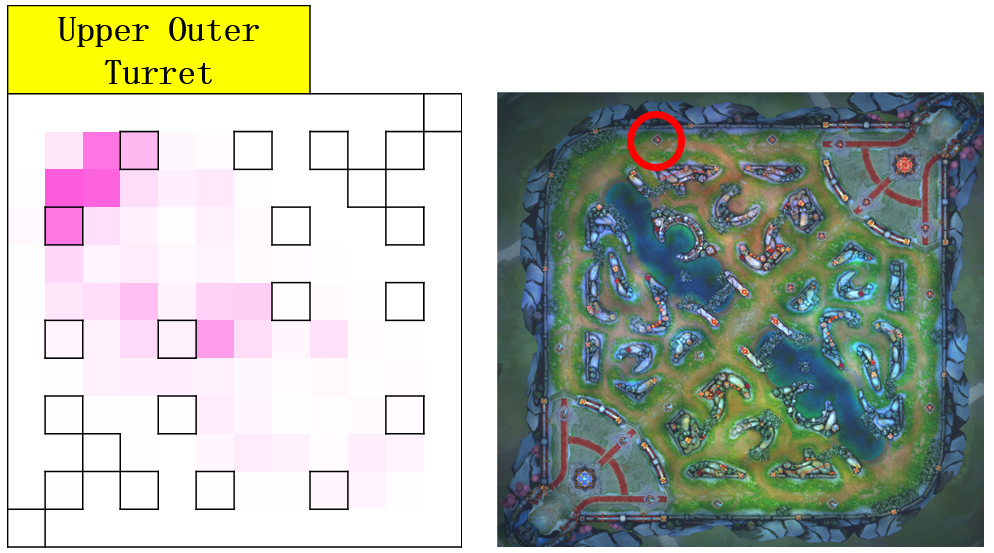}}}}
	\subfloat[]{{{\label{base}\includegraphics[width=0.49\columnwidth]{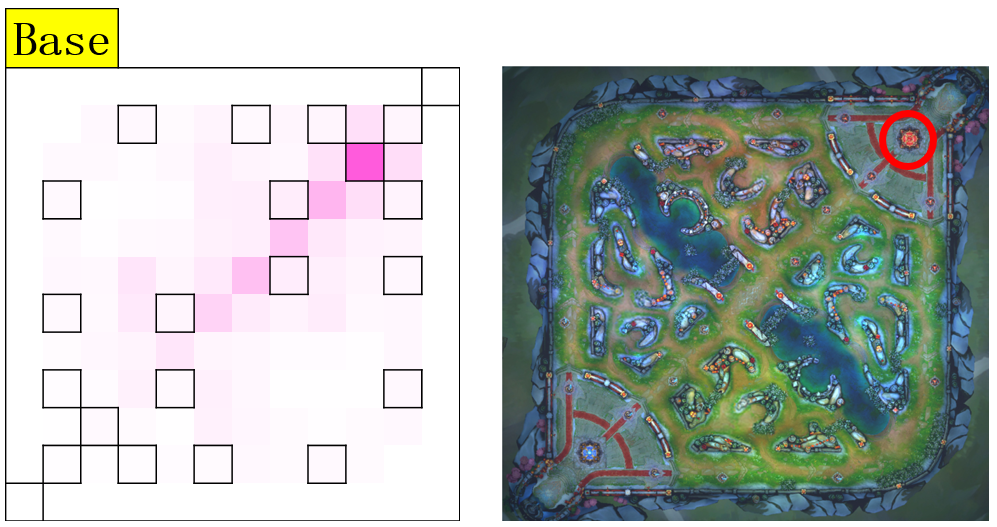}}}}
	\caption{Attention distribution of different strategies. The two attention figures describe attention distribution of the two major resources, i.e., upper outer turret and base respectively.}
	\label{hotspots}
\end{figure}

\begin{figure*}[h]
	\centering
	\includegraphics[width=1.5\columnwidth]{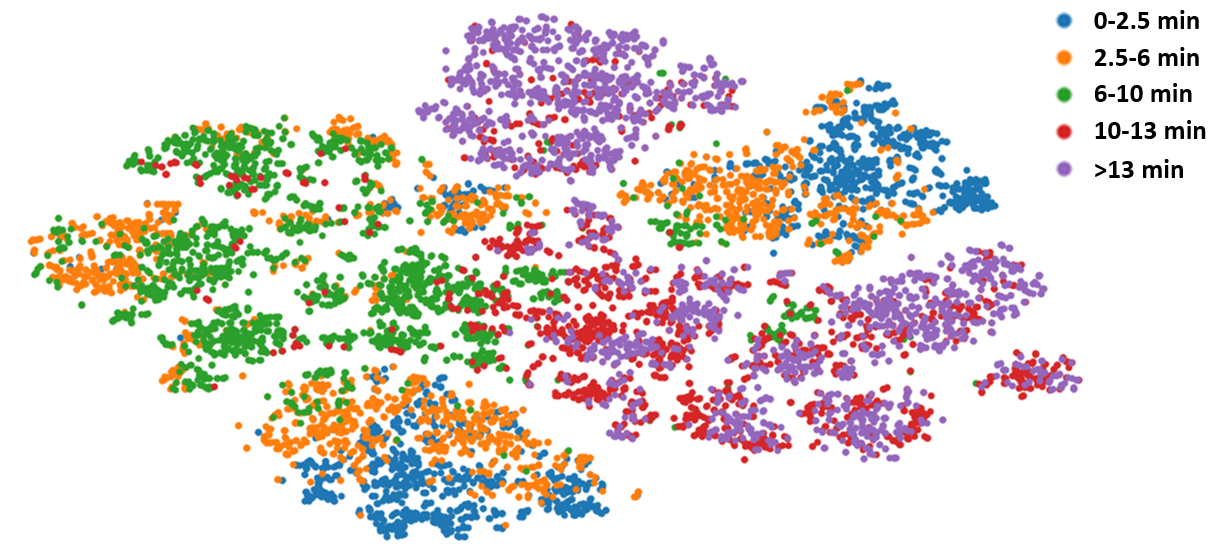}
	\caption{t-Distributed Stochastic Neighbor Embedding on phase layer output. Embedded data samples are coloured with respect to different time stages.}
	\label{tsne}
\end{figure*}

We visualize attention distribution of different phases in Figure. \ref{upper_outer} and \ref{base}. We can see that attention distributes around the major resource of each phase. For example, for upper outer turret phase in Figure. \ref{upper_outer}, the attention distributes around upper outer region, as well as nearby jungle area. Also, as shown in Figure. \ref{base}, attention distributes mainly in the middle lane, especially area in front of the base. These examples show that our phase layer modeling affects attention distribution in practice. To further examine how phase layer correlates with game phases, we conduct t-Distributed Stochastic Neighbor Embedding (t-SNE) on phase layer output. As shown in Figure. \ref{tsne}, samples are coloured with respect to different time stages. We can observe that samples are clearly separable with respect to time stages. For example, blue, orange and green (0-10 minuets) samples place close to one another, while red and purple samples (more than 10 minuets) form another group. 

\subsubsection{Macro Strategy Embedding}

We evaluate how important is the macro strategy modeling. We removed the macro strategy embedding and trained the model using micro level actions from the replays. The micro level model design is similar to OpenAI Five \cite{dota}. Detail description of the micro level modeling is out of the scope of this paper.

The result is listed in Table. \ref{match_stats}, column AI Without Macro Strategy. As the result shows, HMS outperformed AI Without Macro Strategy with 75\% winning rates. HMS performed much better than AI Without Macro Strategy in terms of number of kills, turrets destruction, and gold. The most obvious performance change is that AI Without Macro Strategy mainly focused on nearby targets. Agents did not care much about backing up teammates and pushing lane creeps in relatively large distance. They spent most of the time on killing neutral creeps and nearby lane creeps. The performance change can be observed from the comparison of engagement rate and number of turrets in Table. \ref{match_stats}. This phenomenon may reflect how important macro strategy modeling is to highlight important spots.

\begin{table*}[h]
	\centering
	
	\caption{Match statistics. 250 games were played against Human Teams, while 40 games were played against Without Macro Strategy, Without Communication, and Without Phase Layer, respectively.}
	\begin{tabular}{|L{2.5cm}|L{3.5cm}|L{2.5cm}|L{3.5cm}|L{3.5cm}|}
		\hline
		\hline
		Opponents	&	AI Without Macro Strategy	& Human Teams	&	AI Without Communication	& AI Without Phase Layer\\
		\hline
		Winning rate	&	\textbf{75\%} - 25\%	&	48.3\% - \textbf{51.7\%}	&	\textbf{62.5\%} - 37.5\%	&	\textbf{65\%} - 35\%\\
		\hline
		Kill	&	\textbf{26.0} - 21.1	&	22.6 - \textbf{26.3}	&	\textbf{19.9} - 19.4	&	\textbf{25.6} - 22.8\\
		\hline
		Game Length	&	16.1 min	&	16.1 min	&	18.2 min	&	18.2 min\\
		\hline
		Gold/Min	&	\textbf{2399} - 2287	&	2603 - \textbf{2616}	&	\textbf{2633} - 2554	&	\textbf{2500} - 2333\\
		\hline
		Engagement Rate	&	\textbf{49\%} - 42\%	&	48\% - 48\%	&	\textbf{49\%} - 47\%	&	\textbf{50\%} - 49\%\\
		\hline
		Turrets	&	\textbf{6.1} - 3.2	&	6.1 - \textbf{6.2}	&	\textbf{6.21} - 5.26	&	\textbf{6.73} - 5.42\\
		\hline
		Dragons	&	\textbf{1.22} - 0.2	&	0.55 - 0.55	&	\textbf{0.65} - 0.49	&	\textbf{1} - 0.41\\
		\hline
		Barons	&	\textbf{0.62} - 0.31	&	\textbf{0.64} - 0.61	&	\textbf{0.45} - 0.41	&	\textbf{0.71} - 0.2\\
		\hline
		Dark Barons	&	\textbf{0.41} - 0.22	&	0.36 - \textbf{0.38}	&	0.35 - 0.32	&	\textbf{0.49} - 0.04\\
		\hline
		\hline
	\end{tabular}
	\label{match_stats}
\end{table*}

\subsubsection{Match against Human Players}

To evaluate our AI performance more accurately, we conduct matches between our AI and human players. We invited 250 human player teams whose average ranking is King in Honour of Kings rank system (above 1\% of human players). Following the standard procedure of ranked match in Honour of Kings, we obey ban-pick rules to pick and ban heroes before each match. The ban-pick module was implemented using simple rules. Note that gamecores of Honour of Kings limit commands frequency to a level similar with human.

The overall statistics are listed in Table. \ref{match_stats}, column Human Teams. Our AI achieved 48\% winning rate in the 250 games. The statistics show that our AI team did not have advantage on teamfight over human teams. The number of kills made by AI is about 15\% less than human teams. Other items such as turrets destruction, engagement rate, and gold per minute were similar between AI and human. We have further observed that our AI destroyed 2.5 more turrets than human on average in the first 10 minutes. After 10 minutes, turrets difference shrank due to weaker teamfight ability compared to human teams. Arguably, our AI's macro strategy operation ability is close to or above our human opponents. 

\subsubsection{Imitated Cross-agents Communication}

To evaluate how important the cross-agents communication mechanism is to the AI ability, we conduct matches between HMS and HMS trained without cross-agents communication. The result is listed in Table. \ref{match_stats}, column AI Without Communication. HMS achieved a 62.5\% winning rate over the version without communication. We have observed obvious cross-agents cooperation learned when cross-agents communication was introduced. For example, rate of reasonable opening increased from 22\% to 83\% according to experts' evaluation. 

\subsubsection{Phase layer}

We evaluate how phase layer affects the performance of HMS. We removed the phase layer and compared it with the full version of HMS. The result is listed in Table. \ref{match_stats}, column AI Without phase layer. The result shows that phase layer modeling improved HMS significantly with 65\% winning rate. We have also observed obvious AI ability downgrade when phase layer was removed. For example, agents were no longer accurate about timing when baron first appears, while the full version HMS agents got ready at 2:00 to gain baron as soon as possible.

%

\section{Conclusion and Future Work}

In this paper, we proposed a novel Hierarchical Macro Strategy model which models macro strategy operation for MOBA games. HMS explicitly models agents'  attention on game maps and considers game phase modeling. We also proposed a novel imitated cross-agent communication mechanism which enables agents to cooperate.

We used Honour of Kings as an example of MOBA games to implement and evaluate HMS. We conducted matches between our AI and top 1\% human player teams. Our AI achieves a 48\% winning rate. To the best of our knowledge, our proposed HMS model is the first learning based model that explicitly models macro strategy for MOBA games. HMS used supervised learning to learn macro strategy operation and corresponding micro level execution from high quality replays. A trained HMS model can be further used as an initial policy for reinforcement learning framework.

Our proposed HMS model exhibits a strong potential in MOBA games. It may be generalized to more RTS games with appropriate adaptations. For example, the attention layer modeling may be applicable to StarCraft, where the definition of attention can be extended to more meaningful behaviors such as building operation. Also, Imitated Cross-agents Communication can be used to learn to cooperate. Phase layer modeling is more game-specific. The resource collection procedure in StarCraft is different from that of MOBA, where gold is mined near the base. Therefore, phase layer modeling may require game-specific design for different games. However, the underlying idea to capture game phases can be generalized to Starcraft as well.

HMS may also inspire macro strategy modeling in domains where multiple agents cooperate on a map and historical data is available. For example, in robot soccer, attention layer modeling and Imitated Cross-agents Communication may help robots position and cooperate given parsed soccer recordings.

In the future, we will incorporate planning based on HMS. Planning by MCTS roll-outs in Go has been proven essential to outperform top human players \cite{silver2016mastering}. We expect planning can be essential for RTS games as well, because it may not only be useful for imperfect information gaming but also be crucial to bringing in expected rewards which supervised learning fails to consider.

\bibliography{aaai19}{}

\begin{thebibliography}{}

\bibitem[\protect\citeauthoryear{DeLoura}{2001}]{deloura2001game}
DeLoura, M.~A.
\newblock 2001.
\newblock {\em Game programming gems 2}.
\newblock Cengage learning.

\bibitem[\protect\citeauthoryear{do Nascimento~Silva and
  Chaimowicz}{2015}]{do2015development}
do~Nascimento~Silva, V., and Chaimowicz, L.
\newblock 2015.
\newblock On the development of intelligent agents for moba games.
\newblock In {\em Computer Games and Digital Entertainment (SBGames), 2015 14th
  Brazilian Symposium on},  142--151.
\newblock IEEE.

\bibitem[\protect\citeauthoryear{DOTA2}{2018}]{ti8}
DOTA2.
\newblock 2018.
\newblock The international 2018.
\newblock https://www.dota2.com/international/announcement/.

\bibitem[\protect\citeauthoryear{Foerster \bgroup et al\mbox.\egroup
  }{2016}]{foerster2016learning}
Foerster, J.~N.; Assael, Y.~M.; de~Freitas, N.; and Whiteson, S.
\newblock 2016.
\newblock Learning to communicate to solve riddles with deep distributed
  recurrent q-networks.
\newblock {\em arXiv preprint arXiv:1602.02672}.

\bibitem[\protect\citeauthoryear{Hagelb{\"a}ck and
  Johansson}{2008}]{hagelback2008rise}
Hagelb{\"a}ck, J., and Johansson, S.~J.
\newblock 2008.
\newblock The rise of potential fields in real time strategy bots.
\newblock In {\em Fourth Artificial Intelligence and Interactive Digital
  Entertainment Conference}.
\newblock Stanford University.

\bibitem[\protect\citeauthoryear{Jia \bgroup et al\mbox.\egroup
  }{2014}]{jia2014caffe}
Jia, Y.; Shelhamer, E.; Donahue, J.; Karayev, S.; Long, J.; Girshick, R.;
  Guadarrama, S.; and Darrell, T.
\newblock 2014.
\newblock Caffe: Convolutional architecture for fast feature embedding.
\newblock {\em arXiv preprint arXiv:1408.5093}.

\bibitem[\protect\citeauthoryear{Murphy}{2015}]{moba_hot}
Murphy, M.
\newblock 2015.
\newblock Most played games: November 2015 – fallout 4 and black ops iii
  arise while starcraft ii shines.
\newblock
  http://caas.raptr.com/most-played-games-november-2015-fallout-4-andblack-ops-iii-arise-while-starcraft-ii-shines/.

\bibitem[\protect\citeauthoryear{Ontan{\'o}n and
  Buro}{2015}]{ontanon2015adversarial}
Ontan{\'o}n, S., and Buro, M.
\newblock 2015.
\newblock Adversarial hierarchical-task network planning for complex real-time
  games.
\newblock In {\em Twenty-Fourth International Joint Conference on Artificial
  Intelligence}.

\bibitem[\protect\citeauthoryear{OpenAI}{2018a}]{dota}
OpenAI.
\newblock 2018a.
\newblock Openai blog: Dota 2.
\newblock https://blog.openai.com/dota-2/ (17 Apr 2018).

\bibitem[\protect\citeauthoryear{OpenAI}{2018b}]{openaifive0}
OpenAI.
\newblock 2018b.
\newblock Openai five.
\newblock https://blog.openai.com/openai-five/ (25 Jun 2018).

\bibitem[\protect\citeauthoryear{Schulman \bgroup et al\mbox.\egroup
  }{2017}]{schulman2017proximal}
Schulman, J.; Wolski, F.; Dhariwal, P.; Radford, A.; and Klimov, O.
\newblock 2017.
\newblock Proximal policy optimization algorithms.
\newblock {\em arXiv preprint arXiv:1707.06347}.

\bibitem[\protect\citeauthoryear{Silva and Chaimowicz}{2017}]{silva2017moba}
Silva, V. D.~N., and Chaimowicz, L.
\newblock 2017.
\newblock Moba: a new arena for game ai.
\newblock {\em arXiv preprint arXiv:1705.10443}.

\bibitem[\protect\citeauthoryear{Silver \bgroup et al\mbox.\egroup
  }{2016}]{silver2016mastering}
Silver, D.; Huang, A.; Maddison, C.~J.; Guez, A.; Sifre, L.; Van Den~Driessche,
  G.; Schrittwieser, J.; Antonoglou, I.; Panneershelvam, V.; Lanctot, M.;
  et~al.
\newblock 2016.
\newblock Mastering the game of go with deep neural networks and tree search.
\newblock {\em nature} 529(7587):484--489.

\bibitem[\protect\citeauthoryear{Simonite}{2018}]{openaifive2}
Simonite, T.
\newblock 2018.
\newblock Pro gamers fend off elon musk-backed ai bots—for now.
\newblock https://www.wired.com/story/pro-gamers-fend-off-elon-musks-ai-bots/
  (Aug 23, 2018).

\bibitem[\protect\citeauthoryear{Sukhbaatar, Fergus, and
  others}{2016}]{sukhbaatar2016learning}
Sukhbaatar, S.; Fergus, R.; et~al.
\newblock 2016.
\newblock Learning multiagent communication with backpropagation.
\newblock In {\em Advances in Neural Information Processing Systems},
  2244--2252.

\bibitem[\protect\citeauthoryear{SuperData}{2018}]{hok}
SuperData.
\newblock 2018.
\newblock Worldwide digital games market: February 2018.
\newblock https://www.superdataresearch.com/us-digital-games-market/.

\bibitem[\protect\citeauthoryear{Synnaeve and
  Bessiere}{2011}]{synnaeve2011bayesian}
Synnaeve, G., and Bessiere, P.
\newblock 2011.
\newblock A bayesian model for rts units control applied to starcraft.
\newblock In {\em Computational Intelligence and Games (CIG), 2011 IEEE
  Conference on},  190--196.
\newblock IEEE.

\bibitem[\protect\citeauthoryear{Tian \bgroup et al\mbox.\egroup
  }{2017}]{tian2017elf}
Tian, Y.; Gong, Q.; Shang, W.; Wu, Y.; and Zitnick, C.~L.
\newblock 2017.
\newblock Elf: An extensive, lightweight and flexible research platform for
  real-time strategy games.
\newblock In {\em Advances in Neural Information Processing Systems},
  2656--2666.

\bibitem[\protect\citeauthoryear{Vincent}{2018}]{openaifive}
Vincent, J.
\newblock 2018.
\newblock Humans grab victory in first of three dota 2 matches against openai.
\newblock
  https://www.theverge.com/2018/8/23/17772376/openai-dota-2-pain-game-human-victory-ai
  (Aug 23, 2018).

\bibitem[\protect\citeauthoryear{Vinyals \bgroup et al\mbox.\egroup
  }{2017}]{vinyals2017starcraft}
Vinyals, O.; Ewalds, T.; Bartunov, S.; Georgiev, P.; Vezhnevets, A.~S.; Yeo,
  M.; Makhzani, A.; K{\"u}ttler, H.; Agapiou, J.; Schrittwieser, J.; et~al.
\newblock 2017.
\newblock Starcraft ii: a new challenge for reinforcement learning.
\newblock {\em arXiv preprint arXiv:1708.04782}.

\bibitem[\protect\citeauthoryear{Wender and Watson}{2012}]{wender2012applying}
Wender, S., and Watson, I.
\newblock 2012.
\newblock Applying reinforcement learning to small scale combat in the
  real-time strategy game starcraft: Broodwar.
\newblock In {\em Computational Intelligence and Games (CIG), 2012 IEEE
  Conference on},  402--408.
\newblock IEEE.

\end{thebibliography}
\bibliographystyle{aaai}

\end{document}